\documentclass[a4paper,reprint,aps,pre]{revtex4-2}
\usepackage{graphicx}
\usepackage{dcolumn}
\usepackage{bm}
\usepackage{color}
\usepackage{amssymb}
\usepackage{amsmath}
\usepackage{xfrac}
\usepackage{verbatim}
\usepackage{float}
\usepackage{cleveref}
\usepackage{comment}
\usepackage[utf8]{inputenc}
\usepackage[T1]{fontenc}
\usepackage{url}

\begin{document}

\title{Packing and ejection dynamics of polymers: Role of confinement, polymer stiffness and activity}

\author{Gokul Upadhyay}
\email{gokul@iisermohali.ac.in}
\author{Rajeev Kapri}
\email{rkapri@iisermohali.ac.in}
\author{Anil Kumar Dasanna}
\email{adasanna@iisermohali.ac.in}
\author{Abhishek Chaudhuri}
\email{abhishek@iisermohali.ac.in}
\affiliation{Department of Physical Sciences, Indian Institute of Science Education and Research Mohali, Sector 81, Knowledge City, S. A. S. Nagar, Manauli PO 140306, India}

\date{\today}


\begin{abstract}
 The translocation of biopolymers, such as DNA and proteins, across cellular or nuclear membranes is essential for numerous biological processes. The translocation dynamics are influenced by the properties of the polymers, such as polymer stiffness, and the geometry of the capsid. In our study, we aim to investigate the impact of polymer stiffness, activity, and different capsid geometries on the packing and ejection dynamics of both passive and active polymers. We employ Langevin dynamics simulations for a systematic investigation. We observe that flexible polymers exhibit packing times that are faster than those of their semi-flexible counterparts. 
 Interestingly, for large polymers compared to the capsid size, sphere facilitates faster packing and unpacking compared to ellipsoid, mimicking the cell nucleus and suggesting a geometrical advantage for biopolymer translocation.
 In summary, we observe that increasing activity accelerates both the packing and ejection processes for both flexible and semi-flexible polymers. However, the effect is significantly more pronounced for semi-flexible polymers, highlighting the crucial role of polymer flexibility in these dynamics. These findings deepen our understanding of the intricate interplay between polymer flexibility, capsid geometry, and activity, providing valuable insight into the dynamics of polymer packing and ejection processes.
\end{abstract}

\maketitle 
\section{Introduction}
The organization and dynamics of biological polymers, such as DNA, actin filaments, and microtubules within cells, are crucial for various cellular processes~\cite{alberts2022molecular}. Actin and microtubules, coupled with motor proteins, become active in the presence of ATP and GTP as fuel. This active framework of cross-linked filaments and motor proteins within the confinement of the cell generates mechanical stresses and maintains the structure and dynamics of the cell~\cite{fletcher2010cell,martino2018cellular,aida2012functional,blanchoin2014actin,lau2003microrheology,mackintosh2008nonequilibrium,brangwynne2008cytoplasmic,winkler2020physics}. In addition to cytoskeletal filaments, the processes that involve viral infection by phages~\cite{oppenheim2005switches}, DNA transduction in bacteria~\cite{hartl2009genetics}, and RNA translation~\cite{kohler2007exporting}, require biopolymers to migrate through or operate within confined spaces, such as pores ranging from 1 to 100 nm in size. The packaging of Phage DNA proceeds by the translocation of DNA into a protein shell called the capsid or prohead with the help of a protein enzyme~\cite{fujisawa1997phage}. The translocation of DNA into the prohead is driven by ATP hydrolysis. Thus, all such scenarios are examples of polymers driven out of equilibrium by active processes involving local energy consumption and leading to directed movement or stress~\cite{fang2019nonequilibrium}.

Due to their importance in biological and synthetic systems, single polymer dynamics in an active environment~\cite{harder2014activity,shin2015facilitation,kaiser2014unusual,eisenstecken2016conformational,gupta2019morphological,shee2021semiflexible}, or polymers made up of active monomers~\cite{isele2015self,ghosh2014dynamics,chelakkot2014flagellar,bianco2018globulelike,jiang2014motion,prathyusha2018dynamically,anand2018structure} have been the subject of several theoretical studies in the past decade. They present a minimal system where the competition and effects of thermal, active, and elastic forces can be carefully studied. Despite this large body of literature, the study of active filaments in confinement has not been extensively explored. In confinement, a polymer chain leads to a significant loss in conformational entropy. For DNA encapsulated in a viral capsid, this leads to the accumulation of enormous internal pressure, which viruses or bacteriophages exploit to push DNA out of the phage into the host cytoplasm~\cite{ali2006polymer,ali2008ejection,petrov2013effects,mahalik2013langevin,marenduzzo2009dna,matsuyama2012ejection,milchev2010ejection,park2021effects}. Technologically, the translocation of a polymer through a nanopore has been extensively investigated both experimentally and theoretically~\cite{branton2010potential,wanunu2012nanopores,howorka2009nanopore,keyser2011controlling,sung1996polymer,park1998polymer,muthukumar1999polymer,muthukumar2003polymer,sakaue2007nonequilibrium,sakaue2010sucking,saito2012process,sarabadani2014iso,rowghanian2011force,ikonen2012unifying,ikonen2013influence,ikonen2012influence,bhattacharya2010out,bhattacharya2013translocation,de2010mapping,gauthier2008monte,huang2014conformations,luo2007influence,cohen2011active,caraglio2017driven,jeon2016electrostatic,sarabadani2018theory,suhonen2018dynamics,suma2017pore,suma2020directional,kumar2019complex,plesa2016direct}, due to its potential applications, such as controlled drug delivery, gene therapy, and rapid DNA sequencing. Although such events are necessarily out of equilibrium and nonequilibrium theories such as tension propagation have been successful in describing them, local energy input in active processes leads to complex, unconventional behavior and therefore necessitates further studies.

In fact, numerical investigations have shown that standard blob scaling theories for passive self-avoiding polymers under confinement~\cite{muthukumar1999polymer,muthukumar2001translocation,cacciuto2006self,kantor2004anomalous,sakaue2009dynamics,de2010mapping,linna2014dynamics,polson2018free,polson2019polymer,huang2019scaling,jin2020effect,wang2023behaviors} do not work for active polymers. When restricted within cavities of various geometries, these laws are valid only when the persistence length in the presence of active forces is much smaller than the size of the blob~\cite{das2019deviations}. Further, it has been shown that the ejection dynamics of an active polymer from a cavity is different from that in the presence of an external force field. Although activity provides a driving force in addition to the entropic drive, it also causes the active polymer to collapse, reducing the entropic drive~\cite{li2023ejection}.

For a polymer trying to escape through a cavity opening, it has to move through a space where the rearrangement of the molecules is slow and uneven. Additionally, the persistence length of the polymer makes it difficult to change shape quickly, trapping it within the twists and turns of the cavity. Therefore, the shape of the cavity is very important. It affects how the polymer is arranged when things are steady and how it moves out during ejection. In this study, we show how variations in the sphericity of different capsid shapes affect the speed of packing and ejection of active semiflexible polymers.

\section{MODEL AND METHODS}\label{II}
\subsection{Polymer model}
We model the polymer as a sequence of $N$ beads connected by springs. The consecutive beads interact via the finite extensible non-linear elastic (FENE) potential, given by:
\begin{equation}
	U_{\textrm{bond}}(r)= -\frac{1}{2}kR_0^2\ln{\left(1-\frac{r^2}{R_0^2}\right)},
\end{equation}
where $k$ represents the spring's strength, and $R_0$ is the maximum allowed separation between the connected monomers of the chain. To prevent the polymer from crossing along its length, non-linked beads of the polymer experience excluded volume interactions modeled by the Weeks-Chandler-Anderson (WCA) potential, given by:
\begin{equation}
	U_{\textrm{bead}}(r)=\begin{cases} 4\epsilon\big[\big(\frac{\sigma}{r}\big)^{12}-\big(\frac{\sigma}{r}\big)^6\big]+\epsilon     ,& \text{if } r\leq r_c\\
	0,& \text{if } r > r_c,
	\end{cases}
\end{equation}
where $\sigma$ is the diameter of a bead, and $\epsilon$ defines the strength of the potential. The cutoff distance is given by $r_c=2^{1/6}\sigma$.

The semi-flexibility of the polymer is modeled by introducing a bending potential between consecutive bonds, described as:
\begin{equation}
    U_{\textrm{bend}}(\theta_{i})= \frac{\kappa}{\sigma}(1-\cos{\theta_{i}}),      
    \label{bend}
\end{equation} 
where $\theta_{i}$ is the angle between the $i^{th}$ and $(i-1)^{th}$ bond vectors, and $\kappa$ is the bending modulus of polymer which is related to persistence length $\ell_p$ as $\kappa/k_BT = (d-1)\ell_p/2$, where $d$ is the dimension.

To incorporate activity into our polymer model, we introduce a propulsion force along the polymer chain. This force, which cannot be expressed as a derivative of some potential, models the intrinsic propulsion mechanism found in biological and synthetic active polymers, imparting a directional drive to each segment of the polymer chain as
\begin{equation}
\mathbf{F}_{\text{active}, i} = \frac{f_p}{2}  (\mathbf{\hat{t}}_i+\mathbf{\hat{t}}_{i-1}),
\end{equation}
where $\mathbf{\hat{t}}_i$ and $\mathbf{\hat{t}}_{i+1}$ are $i^{th}$ and $(i+1)^{th}$ unit bond vectors and $f_p$ is the constant magnitude of the active force~\cite{prathyusha2018dynamically}. For the end monomers, only one bond vector will contribute to the expression described above.

This representation allows us to capture the complex behavior of self-propelled polymers, including their interactions with the surrounding environment and their response to external forces or stimuli. The activity is characterized by the dimensionless P\'eclet number (Pe), which for the polymer is defined as

\begin{equation}
Pe = \frac{v_c L}{D_t}=\frac{f_p L^2}{k_b T}.
\end{equation} 
Here, \( v_c = f_p / \gamma_1 \) represents the velocity of the chain contour, and \( D_t = k_B T / (\gamma_1 L) \) denotes the translational diffusion coefficient, where \(\gamma_1\) is the friction per unit length, given by \(\gamma_1 = \zeta (N + 1) / L \) \cite{isele2015self}.

\subsection{Capsid model}
In our simulations, the capsid wall and the capsid pore are represented using monomers that are fixed in space, each with a diameter of $\sigma$. The interaction between the capsid wall and the polymer particles is modeled by the same WCA potential as that used for the polymer. The interaction between the capsid pore and the polymer is attractive, modelled using  the Lennard-Jones (LJ) potential, defined as:
\begin{equation}
    U_{\mathrm{pore}}(r) = 
    \begin{cases} 
        4\epsilon_\text{p} \left[ \left( \frac{\sigma}{r} \right)^{12} - \left( \frac{\sigma}{r} \right)^6 \right], & \text{if } r \leq r_c^{\mathrm{LJ}}, \\
        0, & \text{if } r > r_c^{\mathrm{LJ}},
    \end{cases}
\end{equation}
where, $\epsilon_p$ represents the depth of the potential well, and the cutoff distance is given by $r_c^{\mathrm{LJ}} = 2.5\sigma$.

\begin{table}[b!]
    \centering
    \begin{tabular}{c|c|c|c|c}
        \hline\hline 
        \(a(\sigma)\) & \(c(\sigma)\) & Volume (\(\sigma^3\)) & Surface Area (\(\sigma^2\)) & Sphericity (\(S_p\)) \\
        \hline \hline
        1.40 & 14.05 & 115.37 & 193.37 & 0.59 \\
        1.60 & 10.76 & 115.38 & 170.39 & 0.67 \\
        2.00 & 6.88 & 115.28 & 139.94 & 0.82 \\
        2.50 & 4.41 & 115.45 & 120.61 & 0.95 \\
        2.75 & 3.64 & 115.31 & 116.09 & 0.98 \\
        3.02 & 3.02 & 115.37 & 114.61 & 1.00 \\
        \hline
    \end{tabular}
    \caption{The geometric parameters corresponding to all capsids that are used in the simulations are shown in the table. $a$ and $c$ are minor axis and major axis of ellipsoid respectively. The shapes are chosen such that sphericity varies from $\simeq 0.6$ to $1.0$ and the volumes are kept constant at $\simeq 115~\sigma^3$.}
    \label{tab:ellipsoid_sphericity}
\end{table}

\begin{figure}[ht]
    \includegraphics[width=1.0\linewidth]{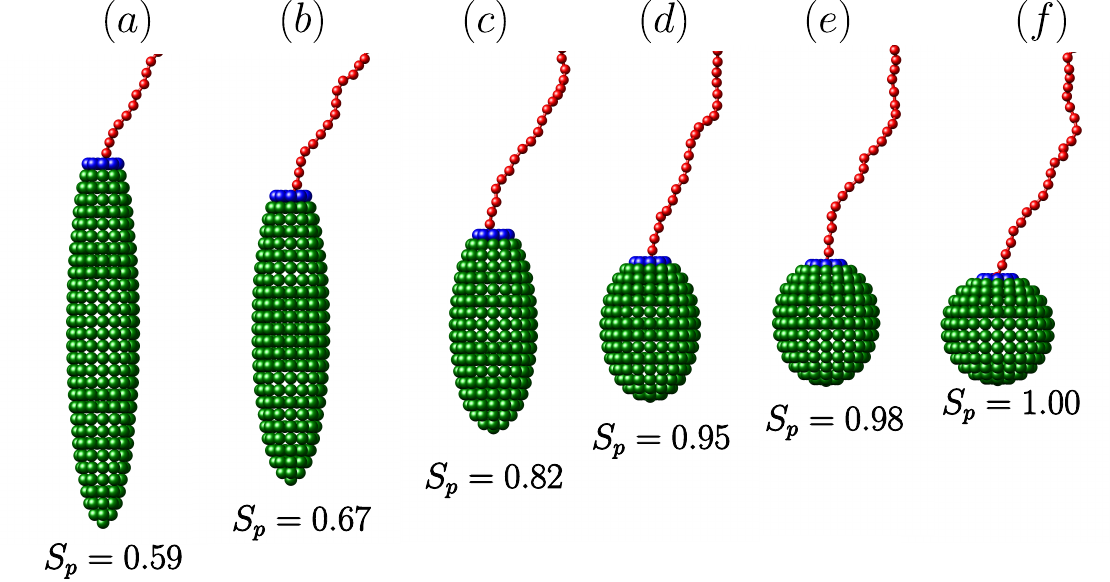}
\caption{ This schematic shows six different capsid geometries, labeled (a) through (f), arranged in increasing order of sphericity. Although all capsids have the same volume, some are depicted in larger form for the sake of clarity. These geometries are used in the packing and unpacking of polymers. Figure (f) represents a perfectly spherical capsid with a sphericity value of 1.}
    \label{capsid-shapes}
\end{figure}

The shape of the capsid is symmetric about the $z-$axis, and the shapes vary from ellipsoidal to spherical. Despite varying the shape of the capsid, we keep the volume of the capsid approximately the same to compare our results. To do so, we measure the sphericity $S_p$, defined as 
\begin{equation}
S_p =\frac{ \pi^{1/3} (6V)^{2/3} }{A},
\end{equation}
where, $V$ is the volume and $A$ is the surface area. Sphericity is a measure of how closely the shape of an object approaches that of a perfect sphere when its value becomes unity. The dimensions and all other details for the various capsid shape are shown in \cref{tab:ellipsoid_sphericity} and plotted in Fig.~\ref{capsid-shapes}.

Although an active polymer can move in and out of the capsid when the active drive is significant, a passive polymer requires an external drive to move inside the capsid. For a comparative study between active and passive polymer packaging and ejection dynamics in various capsid shapes, the polymers experience an additional external force at the entrance of a pore, given by $\textbf{F}_{ext} = F \hat{\textbf{z}}$, with $F=7.0 k_bT/\sigma$. The pore diameter of capsids is taken as $W=2.4\sigma$ to ensure single-file translocation of the polymer during packing and unpacking processes.

\subsection{Simulation details}
We performed all the simulations using LAMMPS, a molecular simulation package ~\cite{plimpton1995fast}, incorporating custom modifications to account for the active forces. The equation of motion for the $i$-th monomer of the polymer is given as : 
\begin{equation}
m\ddot{\textbf{r}}_i = -\boldsymbol{\nabla}U_i+\mathbf{F}_{\text{active}, i}+\textbf{F}_{ext}-\zeta\textbf{v}_i+\boldsymbol{\eta}_i,
\end{equation}
where, $m$ is the mass of the monomer and   $U_i$($=U_{\textrm{bead}}+U_{\textrm{bond}}+U_{\textrm{bend}}+U_{\textrm{wall}}+U_{\textrm{pore}}$) denotes the total potential experienced by a bead, $\zeta$ is the friction coefficient, ${\bf v}_i$ is the monomer velocity and $\boldsymbol{\eta}_i$ is a random force satisfying the fluctuation-dissipation theorem $\langle \boldsymbol{\eta}_i(t)\boldsymbol{\eta}_j(t^{'})\rangle= 6k_bT\zeta\delta_{ij}\delta(t-t^{'})$.

In our model, $\epsilon=k_bT, \sigma$ and $m$ set the units of energy, length, and mass, respectively. This sets the unit of time as $({m{\sigma}^{2}}/{\epsilon})^{1/2}$. Using these units, the dimensionless parameters of $k=30 k_bT/\sigma^2$, $R_0=1.5\sigma$, $k_BT=1$, $\zeta=5 (mk_bT/\sigma^2)^{1/2}$, and $N=100$ have been chosen for the simulations.  Before the packing process started, the first monomer of the polymer was fixed at the entrance of the capsid pore, and the remaining monomers were allowed to equilibrate. The first monomer was then released, and the packing and unpacking dynamics of the polymer was subsequently monitored. In all simulation runs, the time step $\Delta t=0.00002\tau_d$ (where $\tau_d=\sigma^2/D=\sigma^2 \zeta/k_b T$ is the characteristic diffusion time scale) and the averaging were carried out in more than 500 successful packing or unpacking events.


\begin{figure*}[t]
    \includegraphics[width=0.95\linewidth]{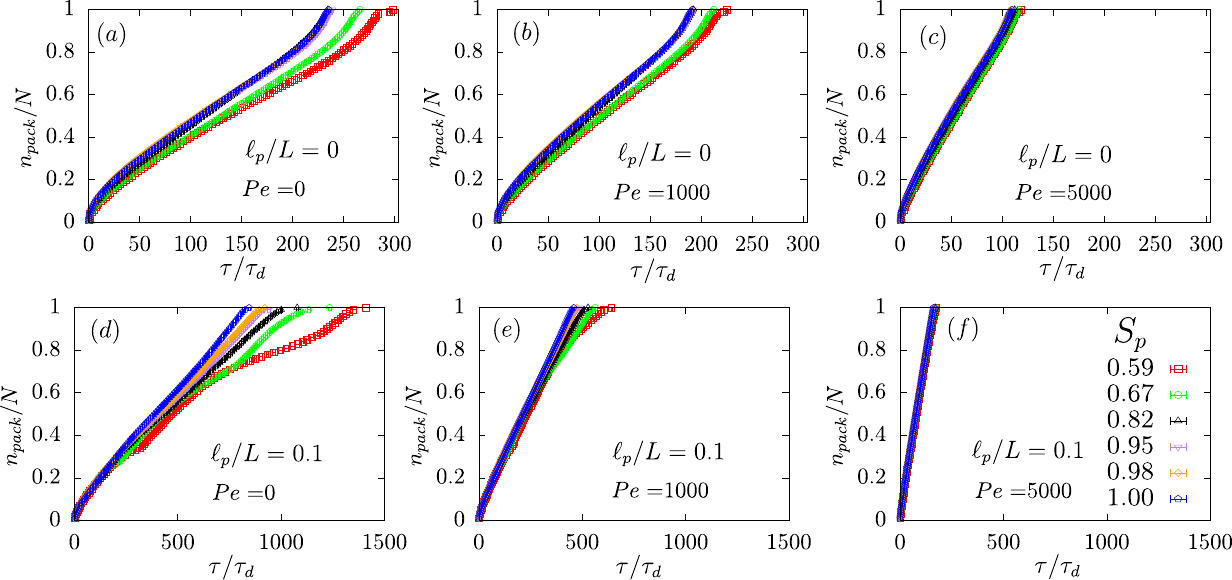}
\caption{ \textbf{PACKING:} Packing times of each monomer for different sphericity values are shown in figures (a) to (c) for $l_p/L=0$ and various $P_e$ values. These figures provide a detailed view of how the packing time of individual monomers changes with increasing $P_e$ under the constraint of $l_p/L=0$. Similarly, figures (d) to (f) illustrate the packing times for $l_p/L=0.1$, allowing a comparison of the packing dynamics between the cases of $l_p/L=0$ and $l_p/L=0.1$, highlighting any differences in the packing behavior of monomers under these two conditions.}
\label{packing_time}.
\end{figure*}

\section{Results}
In the following sections, we study how the polymers are packed into and ejected from the capsids with various geometries. We will examine in detail the effect of semiflexibility and activity in polymer packing into capsids and ejection from capsids with decreasing sphericity. To characterize this, we measure the number of packed or ejected monomers with time and the total packing and ejection time. For all simulations, we fixed the polymer size at $N=100$. Additionally, the first monomer was always placed at the pore in the starting configuration for both packing and ejection to ensure that the entire polymer chain would pack inside or eject out, respectively.
\subsection{Packing}

\subsubsection{Packing dynamics}

In \Cref{packing_time}, the time evolution of the polymer packing is presented. \Cref{packing_time} (a)-(c) show the fraction of packed monomers $n_{pack}/N$ as a function of $\tau/\tau_d$ for flexible polymer (that is, $\ell_p/L = 0$) for various P\'eclet numbers. It is evident from the data that initially the number of packed monomers increases monotonically. The rate of polymer packing slows down once enough polymer is inside the capsid. This happens because the polymer needs to overcome a couple of challenges: overcoming an entropic energetic barrier and finding the optimal arrangement within the capsid. The latter part becomes significant when the capsid's sphericity decreases. Therefore, the overall packing time is highest for the capsid with the lowest sphericity. This dependence on sphericity is most prominent for the passive polymer ($Pe = 0$).  When the polymer becomes active, the overall process accelerates and, at very large P\'eclet numbers, the quantitative differences due to different capsid geometries disappear, as shown in \Cref{packing_time} (c). 

For a semiflexible polymer ($\ell_p/L=0.1$), shown in \Cref{packing_time} (d)-(f), the qualitative behavior remains the same. The stiffness of the polymer provides an additional energy cost during the packing process. A semi-flexible polymer coils inside the capsid, and for shapes with less sphericity, the polymer needs to fold more to pack efficiently. This results in significantly longer packing times for a semiflexible polymer (compare \Cref{packing_time} (a) and (d)). Again, as expected, the difference in packing times within capsids of various sphericity reduces with increasing activity (compare \Cref{packing_time} (e)-(f)).

\begin{figure}[h]
    \includegraphics[width=0.75\linewidth]{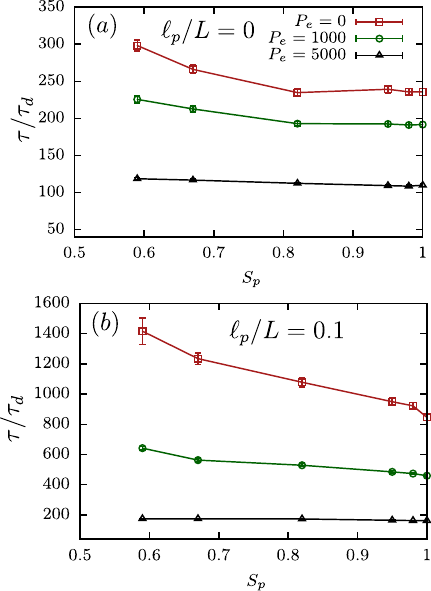}
\caption{\textbf{PACKING:} Total packing time as a function of sphericity $S_p$ for different P$\acute{\text{e}}$clet numbers.(a) corresponds to $\ell_p/L=0$ and (b) corresponds to $\ell_p/L=0.1$.
}
\label{Total_packing}
\end{figure}

\begin{figure}[b]
    \includegraphics[width=0.75\linewidth]{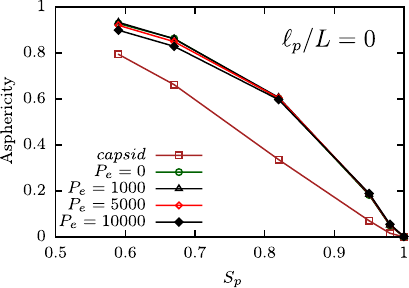}
\caption{The shape of the polymer inside different confinements during both packing and unpacking for various P$\acute{\text{e}}$clet numbers. It also illustrates the capsid shapes based on their sphericity.}
\label{asphericity}
\end{figure}
\subsubsection{Total Packing times}

\begin{figure*}[t]
    \includegraphics[width=0.9\linewidth]{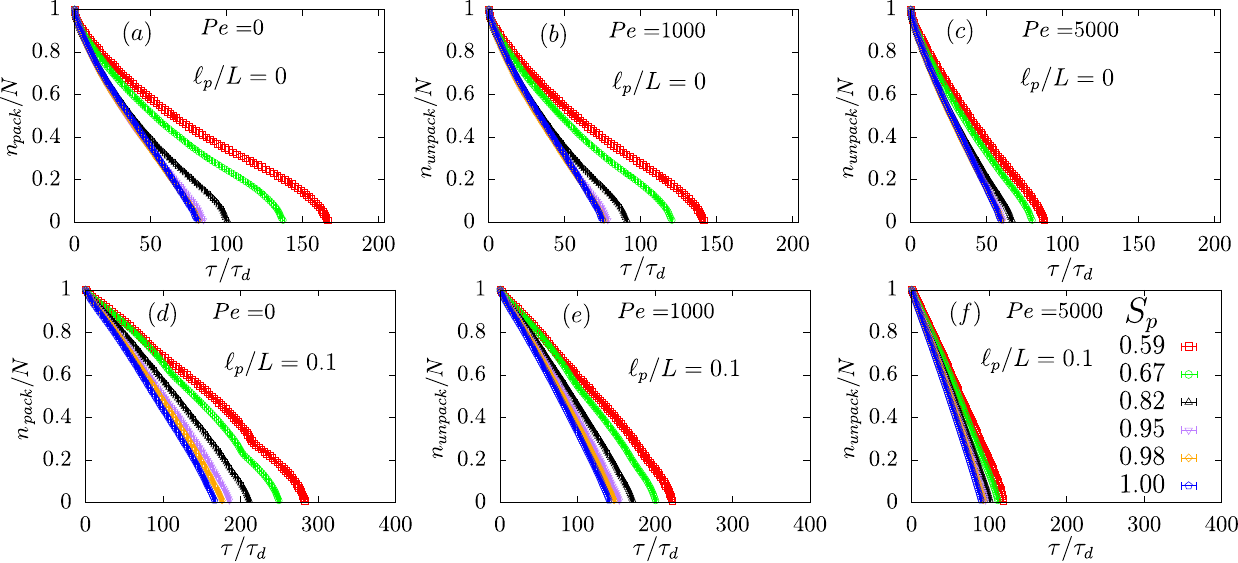}
\caption{\textbf{EJECTION:} (a) to (c) display the ejection times of each monomer for different sphericity values at  $l_p/L=0$ and various $P_e$ values, providing a detailed view of how the ejection time changes with increasing  $P_e$ under the constraint of $\ell_p/L=0$. Figures (d) to (f) illustrate the ejection times for $\ell_p/L=0.1$, allowing a comparison of the ejection dynamics between $\ell_p/L=0$ and $\ell_p/L=0.1$, highlighting differences in the ejection behavior of monomers under these two conditions.}
\label{ejection_time}.
\end{figure*}

\begin{figure}[h]
    \includegraphics[width=0.75\linewidth]{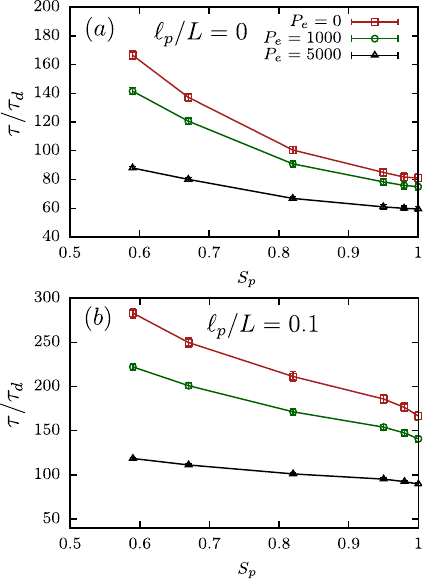}
\caption{ \textbf{EJECTION:} Total ejection time as a function of sphericity $S_p$ for different P$\acute{\text{e}}$clet numbers. Figure (a) illustrates the ejection times when $l_p/L=0$, highlighting the impact of varying sphericity on the total ejection time for flexible polymers. Figure (b) presents the ejection times for $l_p/L=0.1$, showing the corresponding effect on semi-flexible polymers. These comparisons emphasize how the polymer's flexibility and the capsid's sphericity influence the overall ejection dynamics across different P$\acute{\text{e}}$clet numbers.}
\label{Total_ejection}
\end{figure}

\begin{figure*}[hbt]
  \centering
    \includegraphics[width=0.9\linewidth]{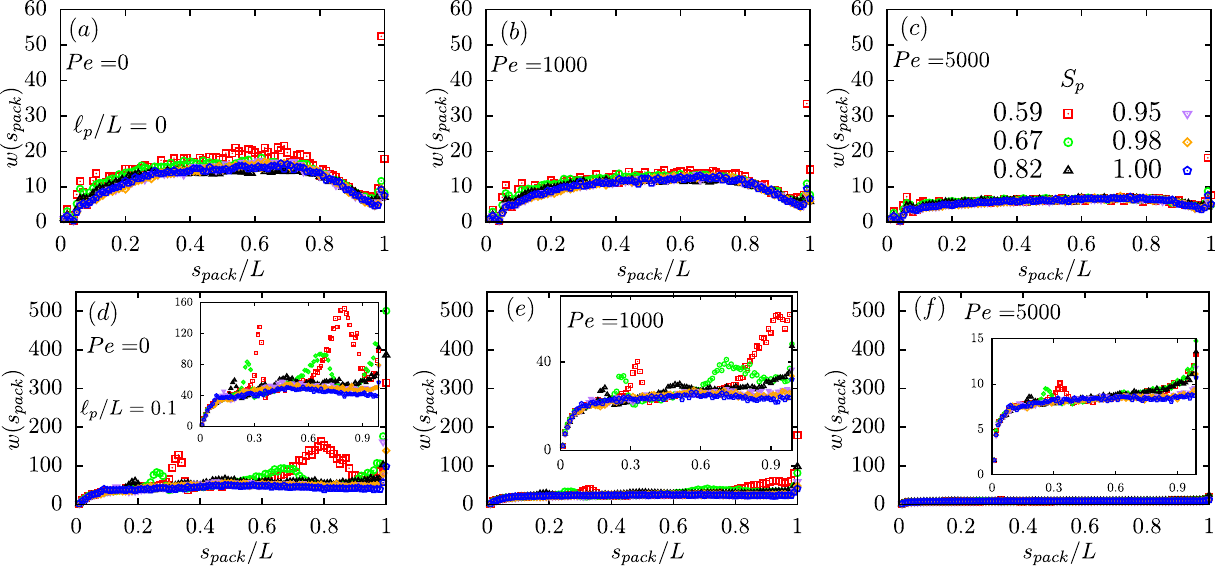}
\caption{\textbf{PACKING:}(a)–(c) Mean waiting times \(w(s_{pack})\) for monomers of a flexible polymer, and (d)–(f) for a semiflexible polymer, at different sphericities $S_p$ and different P\'eclet numbers $P_e$. In (a), we can see the effect of tension propagation. However, as we increase the P\'eclet number or increase the rigidity, this effect diminishes. For the semiflexible polymer, we observe more prominent oscillation peaks during packing at low sphericity, indicating the polymer's unfolding process. }
\label{WT_Total_packing}
\end{figure*}

In \Cref{Total_packing} (a), we have shown the total packing time against capsid sphericity for various P\'eclet numbers. As mentioned in the previous section, the total packing time decreases with increasing sphericity. Above a sphericity of $0.8$, the total packing time shows minimal variation, indicating a negligible influence of the shape of the capsid beyond this value.  
As we increase $Pe$, the qualitative behavior remains the same, but the difference in the packing time between different sphericities continuously decreases. Therefore, provided the volume of the cavity is the same, the shape of the cavity does not influence the total packing times of an active polymer at high $Pe$.  In \Cref{Total_packing} (b), at $\ell_p/L=0.1$ the packing time increases for all sphericity and for all P\'eclet numbers compared to the case $\ell_p/L=0$, as expected. 

To understand whether packed monomers occupy the entire volume, we measured the asphericity of the completely packed polymer for various P\'eclet numbers. To measure the asphericity, we first calculate the gyration tensor \(\mathbf{G}\) which is defined as:
\begin{equation}
    G_{\alpha\beta} = \frac{1}{N} \sum_{i=1}^{N} (r_{i,\alpha} - R_{\alpha})(r_{i,\beta} - R_{\beta}),
    \label{eq:gyration}
\end{equation}
where, \(N\) is the number of monomers, \(r_{i,\alpha}\) is the $\alpha$th coordinate ($\alpha \equiv (x, y, z)$) of the $i$th monomer and \(R_{\alpha}\) is the $\alpha$th coordinate of the center of mass, defined as: $R_{\alpha} = (1/N) \sum_{i=1}^{N} r_{i,\alpha}$. From the gyration tensor, we can obtain the asphericity~\cite{rudnick1987shapes} of the polymer configuration as:
\begin{equation}
    A_{sp} = \frac{ \left( \lambda_1 - \lambda_2 \right)^2 + \left( \lambda_2 - \lambda_3 \right)^2 + \left( \lambda_3 - \lambda_1 \right)^2 }{2 R_g^4 },
    \label{eq:asphericity}
\end{equation}
where, $\lambda_1$, $\lambda_2$, and $\lambda_3$ are the eigenvalues of the gyration tensor and $R_g^2 = (\lambda_1^2 + \lambda_2^2 + \lambda_3^2)/N$.

The asphericity is plotted against the capsid sphericity and shown along with capsid's asphericity in \Cref{asphericity} for a flexible polymer ($\ell_p/L =0$).  Packed polymers mimic the capsid's asphericity, conforming to the overall shape during packing. The slight difference arises from the capsid's smaller internal dimensions compared to its external size, causing the polymer to exhibit a slightly higher asphericity upon packing. We have also checked for a semiflexible polymer ($\ell_p/L =0.1$) and found a similar behavior.

\subsection{Ejection}

\subsubsection{Ejection dynamics}

In \Cref{ejection_time}, the time evolution of the number of monomers that have escaped the capsid is shown for various capsid geometries and P\'eclet numbers. From \Cref{ejection_time} (a)-(c), it is clear that as the sphericity of the capsid decreases, the time it takes for a fraction of monomers to escape increases.  However, this effect weakens at very high values of P\'eclet number. Note that the initial unpacking/ejection happens fast, with a sharp decrease in the ejection time for all sphericities. However, the slope becomes smaller for end fractions of the polymer that escapes the pore. There are two reasons for this. First, the polymer that is packed inside the pore needs to unfold. Secondly, the attractive interactions at the pores pull on the polymer leading to longer waiting times for the end beads. 

For a semiflexible polymer ($\ell_p/L=0.1$), this qualitative behavior remains consistent (\Cref{ejection_time} (d)-(f)). The ejection times for smaller values of P\'eclet are higher than for a flexible polymer. One feature in the ejection times is the appearance of kinks, suggesting that the escape process of the semiflexible polymer is in a jerky motion.  With increasing P\'eclet, this effect becomes less dominant. We shall discuss this jerky motion later. 

\subsubsection{Total ejection times}
As evidenced in the last section, the total ejection time of a polymer for different activity levels is significantly influenced by the shape of the confinements. We look at the scaled total ejection time $\tau/\tau_d$  of the polymer as a function of the capsid's sphericity for different $P_e$ values, considering both flexible and semi-flexible polymers. In \Cref{Total_ejection} (a), we present the results for $\ell_p/L=0$ and in (b) for $\ell_p/L=0.1$. For a passive polymer ($P_e=0$), we observe that as the sphericity of the capsid increases, the total ejection time decreases monotonically, indicating that a spherical capsid is more favorable for polymer ejection, as we observed in packing. Also, as in packing, increasing activity speeds up the ejection process. However, semi-flexible polymers take longer to exit the capsid than flexible ones.

\subsection{Waiting time}

Next, we calculate the mean waiting time of each monomer, $ w(s_{pack})$, as it packs and unpacks within the confinement for various P\'eclet numbers and for different capsid geometries. The mean waiting time of a monomer is obtained by calculating the average time it spends inside the pore. It is an important quantity that reveals the dynamics of the translocation process at the monomer level. In ~\Cref{WT_Total_packing}(a)-(c) the mean waiting times \(w(s_{pack})\) for a flexible polymer are shown, while ~\Cref{WT_Total_packing}(d)-(f) represent the same for a semi-flexible polymer, each at various sphericities \(S_p\) and P\'eclet numbers \(P_e\). The waiting time dynamics can be understood using the concept of tension propagation along the backbone of the polymer ~\cite{sakaue2007nonequilibrium,sakaue2010sucking,saito2011dynamical,dubbeldam2012forced,ikonen2012unifying,ikonen2012influence}. When a polymer is pulled through a pore by an external driving force, tension propagates along the polymer backbone. This driving force acts on the beads inside the pore and facilitates entry and exit, especially at low P\'eclet numbers. A tension front divides the moving and nonmoving sections of the polymer chain. The effective drag caused by the moving part of the chain increases over time as more monomers experience tension. This results in longer waiting times for the initial monomers. Once the tension front reaches the end monomer, the tension propagation process stops and the entire chain is pulled toward the pore, leading to shorter waiting times for the monomers. This explains the initial rise, flattening and then drop of $w(s)$ with monomer number. The rise in $w(s)$ for the end monomers is a consequence of the attractive interaction at the pore that holds back the end monomers. An additional active drive on the polymer backbone leads to shorter waiting times of the individual monomers and very little influence of the attractive nature of the pores on the end ones. For the flexible polymer, sphericity plays an important role at low activity. This is entirely consistent with longer packing/ejection times for lower sphericities.

For the packing of a semiflexible polymer, we observed prominent peaks in the mean waiting times for capsid geometries with low sphericities
(see \Cref{WT_Total_packing}(d)-(f)). As the semiflexible polymer packs inside the capsid, the geometric constraint coupled to the energy cost of bending implies that the polymer segment inside the capsid needs to coil adequately to allow the remaining part of the chain to enter. This appears as bursts in the mean waiting time when the capsid shapes are more elongated. These monomers have to wait longer to allow the polymer to fold inside the capsid at its minimum energy/maximum entropy conformation.  

\begin{figure*}[ht]
  \centering
    \includegraphics[width=0.9\linewidth]{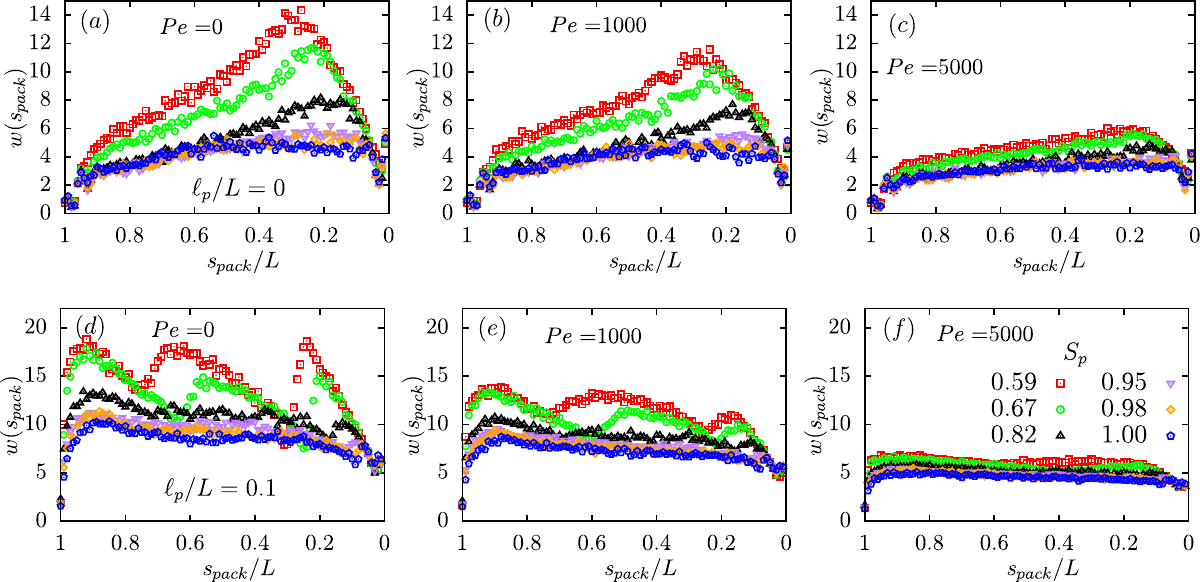}
\caption{ \textbf{EJECTION:} (a)–(c) show the average waiting times \(w(s_{pack})\) for monomers in a flexible polymer, while (d)–(f) display the same for a semiflexible polymer, both at varying sphericities \(S_p\) and P$\acute{\text{e}}$clet numbers \(P_e\). In (a), we observe the influence of tension propagation, which lessens as the P$\acute{\text{e}}$clet number increases. For the semiflexible polymer, we notice more prominent oscillation peaks during unpacking at low sphericity, indicating the polymer's unfolding process.}
\label{WT_Total_ejection}
\end{figure*}

In the ejection process, the dynamics observed in the mean waiting times of monomers are even more striking (see \Cref{WT_Total_ejection}). For a flexible polymer, the peak in mean waiting time is robust for low activities, in sharp agreement with the tension propagation theory. Because polymer ejection from the capsid is an entropically favorable process, the role of attractive interactions at the pore does not affect the dynamics of the end monomers significantly. The role of sphericity is visible in the waiting time characteristics as the curves are distinctly separated, and the waiting time increases for all monomers with the lowering of sphericity. 

\begin{figure}[h]
  \centering
\includegraphics[width=0.85\linewidth]{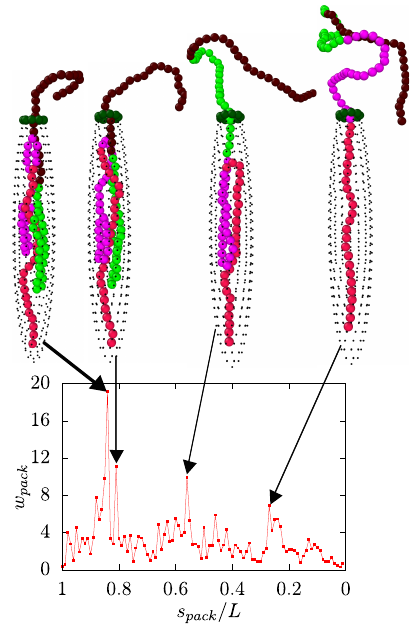}
\caption{\textbf{EJECTION:} Data from one simulation of a semiflexible polymer ejecting from a capsid with a sphericity of 0.59. The snapshots capture different stages of the polymer folding inside the capsid, shown at the peak waiting time.}
\label{single_ejection_time}.
\end{figure}
For the semiflexible polymer, the waiting time characteristics now show prominent oscillations at low sphericity and smaller activity. We claim that this behavior is again related to the combined interplay of confinement and bending rigidity of the polymer. To substantiate, we show
snapshots of a typical polymer ejection from a low sphericity capsid in \Cref{single_ejection_time}. In this plot, the waiting time is displayed in relation to the polymer contour distance. Polymer snapshots at various stages of ejection are also shown. To provide visual clarity of the folds, we have shown the polymer sections in four different colors. From the data, it is clear that each peak is associated with an unfolding event. The unfolding events are, of course, dependent on the capsid cross section and bending rigidity of the polymer.

\section{Discussions}
Our study highlights the significant impact of the capsid shape on the packing and ejection of semi-flexible filaments. Using Langevin dynamics, we showed that spherical capsids allow faster packing and unpacking compared to ellipsoidal ones. This is because spherical shapes reduce internal friction and provide easier paths for the polymers to move. In ellipsoidal capsids, 
polymers go through multiple folds which make both packing and unpacking a slow process. We also showed that flexible polymers packed faster than semi-flexible polymers as a result of their higher adaptability and lower resistance to compression. We found that activity played a passive role in both packing and unpacking, such as, increased activity levels reduced the packing time further, especially for semi-flexible polymers, by enhancing their movement and shape changes. 
\\
Interestingly, in case of spherical capsid both flexible and semi-flexible polymers exhibit almost similar total ejection times, but packing times differ significantly. Semi-flexible polymers require considerably more time to pack compared to their flexible counterparts. In \cite{ali2006polymer}, authors also showed that there is a difference between flexible and semi-flexible polymers for both ejection and packing process. However, in our study the difference between flexible and semi-flexible polymer packing times is substantial for all capsid geometries. 
With regard to active polymers, irrespective of whether flexible or semi-flexible, both the ejection and the packing dynamics speed up. This has also been observed recently in case of ejection of flexible polymers~\cite{li2023ejection}, where the authors found that the total ejection time scales as $\tau_\text{ej}\simeq \text{Pe}^{-0.97}$ for flexible polymers. It appears that for this system the activity merely speeds up the whole process rather than coupling with polymer elasticity giving rise to interesting dynamics.

While our simplified model using Langevin simulations effectively demonstrated the influence of polymer semi-flexibility, capsid geometry, and polymer activity on packing and ejection, it lacks several details that present in real biological systems. Notably, both the pore and capsid are deformable, which would impact the dynamics of these processes. For instance capsid can be modelled as triangulated surface~\cite{dasanna2024mesoscopic} instead of rigid surface. Capturing those details necessitates more detailed models, a topic we leave for future investigations.

\newpage
\bibliographystyle{apsrev4-2}
\bibliography{bibfile.bib}

\end{document}